\documentclass[a4paper,twocolumn,11pt,tightenlines,superscriptaddress,showkeys,showpacs]{revtex4-1}

\usepackage{graphicx}
\usepackage{dcolumn}
\usepackage{bm}
\usepackage{xcolor}
\usepackage{xspace}

\newcommand*{\pT}{\ensuremath{p_\mathrm{T}}\xspace}

\newcommand*{\kT}{\ensuremath{k_\mathrm{T}}\xspace}
\newcommand*{\pTjet}{\ensuremath{p_\mathrm{T}^{\mathrm{jet}}}\xspace}

\begin{document}


\title{KNO-like scaling within a jet in proton--proton collisions at LHC energies}

\author{R\'obert V\'ertesi}
\email{vertesi.robert@wigner.hu}
\affiliation{Wigner Research Centre for Physics, MTA Centre of Excellence,\\ 
	29-33 Konkoly-Thege Mikl\'os str., 1121 Budapest, Hungary}

\author{Antal G\'emes}
\affiliation{Trinity College, University of Cambridge, Cambridge CB2 1TQ, United Kingdom}
\affiliation{Wigner Research Centre for Physics, MTA Centre of Excellence,\\ 
	29-33 Konkoly-Thege Mikl\'os str., 1121 Budapest, Hungary}


\author{Gergely G\'abor Barnaf\"oldi}
\affiliation{Wigner Research Centre for Physics, MTA Centre of Excellence,\\ 
	29-33 Konkoly-Thege Mikl\'os str., 1121 Budapest, Hungary}

\date{\today}

\begin{abstract}
We study the multiplicity distributions of events with hard jets in proton-proton collisions at LHC energies using \textsc{Pythia} 8 Monte-Carlo simulations. We demonstrate that the charged-hadron multiplicity distributions scale with jet momentum. This suggests that the Koba--Nielsen--Olesen (KNO) scaling holds within a jet. The in-jet scaling is fulfilled without multiple-parton interactions (MPI), but breaks down in case MPI is present without color reconnection. Our findings imply that KNO scaling is violated by parton shower or multiple-parton interactions in higher-energy collisions. 
\end{abstract}

\maketitle
{\it\label{sec:intro}Introduction. --- }
Final-state multiplicities in high-energy collisions are known to follow a negative binomial distribution (NBD)~\cite{ALICE:2017pcy}.
It has been demonstrated by Koba, Nielsen and Olesen that multiplicity distributions at different collision energies $\sqrt{s}$ collapse to a single distribution using a simple scaling (the KNO scaling)~\cite{Koba:1972ng,Landua:1970wz}. It was later found, however, that the KNO scaling breaks down at higher $\sqrt{s}$~\cite{Arneodo:1987qy,Alner:1985wj}. While several explanations have been proposed~\cite{Hegyi:1996wb,Kudo:1986mk,Burgers:1987za,Matinyan:1998ja}, a complete understanding of the origin of the scaling and its violation is missing up to this day~\cite{Hegyi:2000sp}.
At higher center-of-mass energies, where the average final-state multiplicity is higher, semi-hard vacuum-QCD effects such as multiple-parton interactions (MPI) play a non-negligible role. It has been assumed in multiple works that scaling violation may be caused by these effects. A scenario based on the Lund string model~\cite{Abramovsky:2007ks,Abramovsky:1980yc} proposes that overlapping color strings break the scaling, while another work argues that underlying-event activity linked to MPI with color reconnection (CR) is responsible for the violation of the scaling~\cite{Ortiz:2018vgc}.
Earlier works propose that the KNO scaling may be a property of the jet itself~\cite{Bassetto:1987fq}. At lower collision energies, since events with jet events have very little background, collision energy can be directly linked to average jet-momentum (\pTjet). This suggests that \pTjet may be a more fundamental scaling variable, and the violation may be explained by the breaking-down of the connection of $\sqrt{s}$ to the 
average \pTjet toward higher energies as well as more complex colliding systems (such as p--A systems examined in Ref.~\cite{Dumitru:2012yr}).

Recently, jet structures in high-energy pp collisions have been found to exhibit a multiplicity-dependent characteristic size, largely independently of the event generator settings and fragmentation functions~\cite{Varga:2018isd,Varga:2019rhi}, that can be explained by scaling properties of the radial jet profiles with multiplicity in any given \pTjet window~\cite{Gemes:2020cfa}. These two scaling properties may be a result of the same statistical processes that govern jet fragmentation~\cite{Takacs:2018fqk}, such as temperature fluctuations as outlined in Refs.~\cite{Wilk:2012zn,Biro:2014yoa}.

In the followings we utilize full event simulations to examine the question whether KNO scaling is restored if a single jet is considered.
The results below can help in understanding the origin of KNO scaling and find the mechanism responsible for its violation.
\begin{figure*}[t]
	\centering
	\includegraphics[width=0.5\textwidth,keepaspectratio]{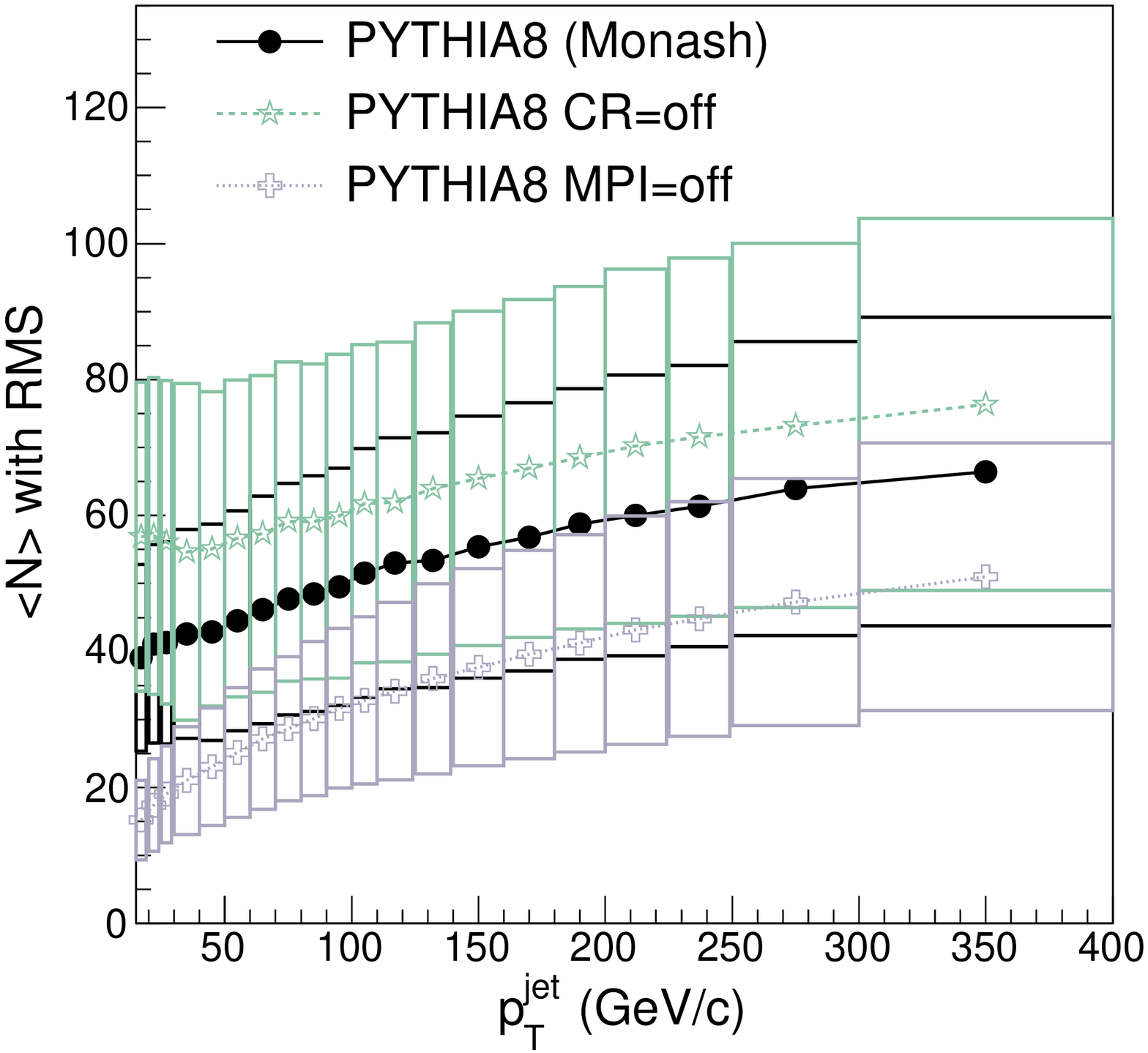}%
	\includegraphics[width=0.5\textwidth,keepaspectratio]{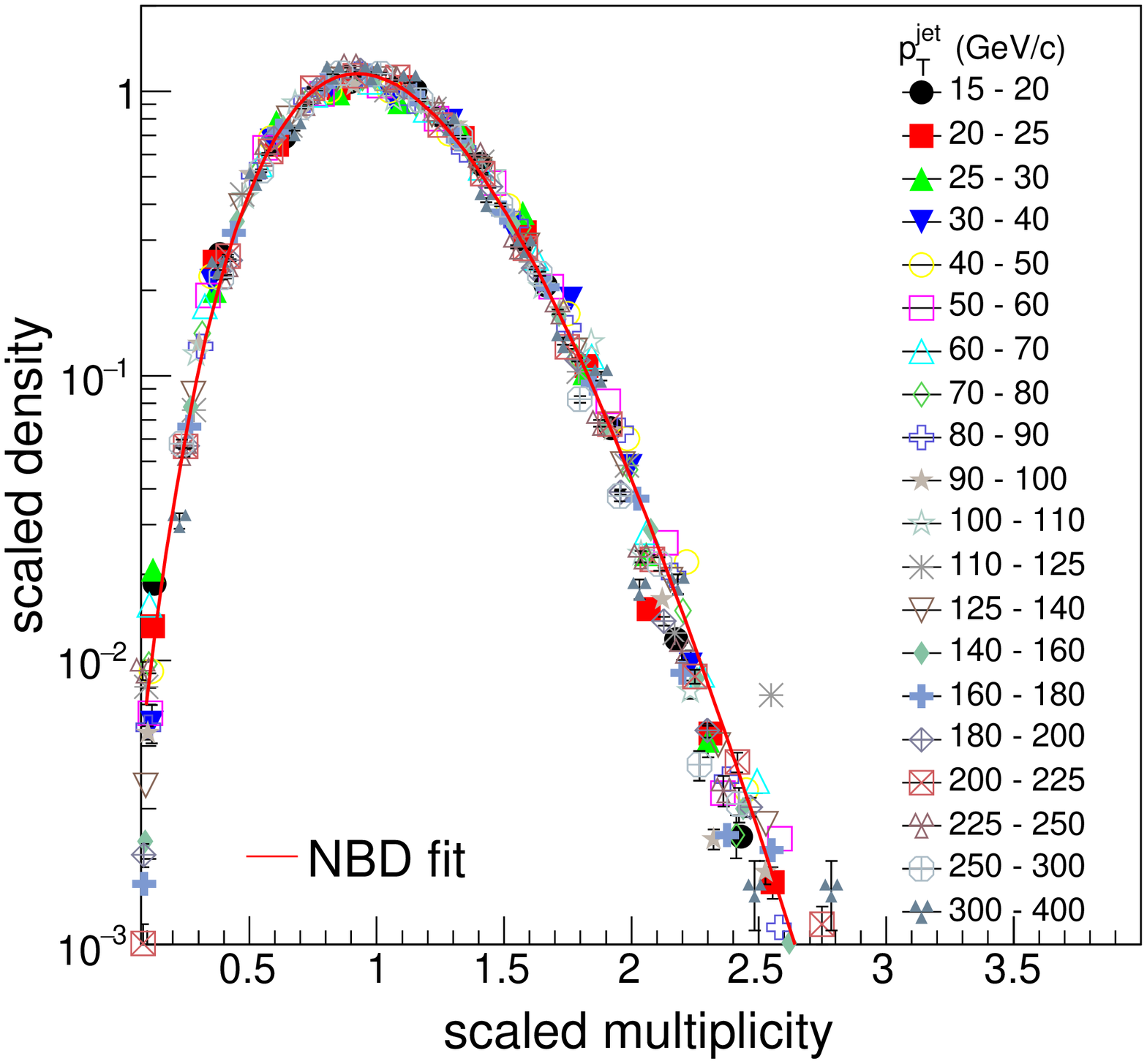}%
	\begin{center}
		\caption{\label{fig:multdist} (Left:) Means of the multiplicity distributions, together with their RMS indicated as boxes, in function of \pTjet for the Monash physical settings as well as for the scenarios without CR and MPI. (Right:) Multiplicity distributions in all \pTjet windows scaled by their means, with an NBD fit.}
	\end{center}
\end{figure*}
%

{\it\label{sec:ana}Analysis method. --- }
Events with high-momentum jets were simulated following the procedure described in previous works~\cite{Varga:2018isd,Gemes:2020cfa}. We used a four samples of 5 million proton-proton collisions at $\sqrt{s}=7$ TeV collision energy, obtained using \textsc{Pythia} 8~\cite{Sjostrand:2014zea} with the Monash tune~\cite{Skands:2014pea}. In the four samples, the minimum momentum transfer of the leading hard scattering, $\hat{\pT}$, was restricted to remain above 5, 20, 40 and 80 GeV/$c$ respectively, to enhance statistics.
Identified pions, kaons and (anti)protons with a lower momentum threshold $\pT>150$ MeV/$c$ were considered as charged hadrons in this analysis. We reconstructed jets using the anti-$\kT$ clustering algorithm~\cite{Cacciari:2008gp} with a resolution parameter $R=0.7$ in the mid-rapidity range $|\eta|<1$. 
Jets were reconstructed in 20 \pTjet windows from 15 up to 400 GeV/$c$ from one of the four datasets the $\hat{\pT}$ chosen so that statistics is maximized without biasing the sample~\cite{Varga:2018isd}.
Simulations were repeated with different settings, where MPI or CR were turned off. (Note that CR is part of the MPI framework, therefore no MPI means no CR either.)

\textsc{Pythia} is extensively tuned to describe the fundamental physical observables of the leading hard process as well as the underlying event, and it is known to reproduce final-state multiplicities well~\cite{Chatrchyan:2013ala,Sirunyan:2017zmn}. 
Since final-state multiplicities in data containing hard processes are dominated by the multiplicities within the jets, we parameterize the behavior with event multiplicity in this study. We examine the scaling of the multiplicity with respect to \pTjet to understand whether, and to what extent, the scaling behavior can be observed.
We have also repeated our studies using simulations with tune-4C~\cite{Corke:2010yf} as well as the MonashStar~\cite{Khachatryan:2015pea} tunes, and the \kT~\cite{Catani:1993hr,Ellis:1993tq} and the Cambridge--Aachen jet clustering algorithms~\cite{Dokshitzer:1997in,Wobisch:1998wt} to rule out the possibility that our findings are a peculiarity of the jet definition, given sets of simulation settings or parton distribution functions.

{\it\label{sec:result}Results. --- }
Figure~\ref{fig:multdist} (left) shows the mean event multiplicities and their RMS in function of \pTjet for the physical settings as well as when either MPI or CR is turned off.
As expected, an increase can be observed towards higher \pTjet. Using this mean, the different distributions can be collapsed onto each other. This is demonstrated on Fig.~\ref{fig:multdist} (right), together with a negative binomial distribution (NBD) fit on all the points in the form
\begin{equation}
P_N=\frac{\Gamma(Nk+a)}{\Gamma(a)\Gamma(Nk+1)}p^{Nk}(1-p)^a\, ,
\end{equation}
where $a, k$ and $p$ are parameters related to the mean and dispersion of the distribution of the multiplicity $N$. The fits to the NBD curve are statistically good ($\chi^2/{\rm NDF}<8$ in each case), suggesting that a KNO-like scaling of the multiplicities with \pTjet instead of $\sqrt{s}$ may be valid in pp collisions at LHC energies. In order to mitigate the effect of fluctuations and quantify this above statment, however, it is useful to have a look at higher moments of the distribution.

The moment of the $q^{\rm th}$ order of the multiplicity distributions in a given $\pTjet$ window can be defined as
\begin{equation}
\left<N^q\right> = \sum_{N=1}^{\infty}P_N N^q
\end{equation}
where $N$ is the charged-hadron event multiplicity and $P_N$ is the probability distribution of the multiplicity.
In case the scaling is fulfilled and the mean scales with a factor $\lambda$, it is expected that the $q^{\rm th}$ moment scales with $\lambda^q$, that is, 
\begin{equation}
\label{eq:lambda}
\left<N^q(\pTjet)\right> = \lambda^q(\pTjet) \left<N^q(p_0)\right>
\end{equation}
where $p_0$ is chosen so that $\lambda(p_0)=1$, and the 
$\lambda^q(\pTjet)$ function otherwise follows the shape shown in Fig.~\ref{fig:multdist} (left).

\begin{figure}[t]
	\centering
	\includegraphics[width=0.5\textwidth,keepaspectratio]{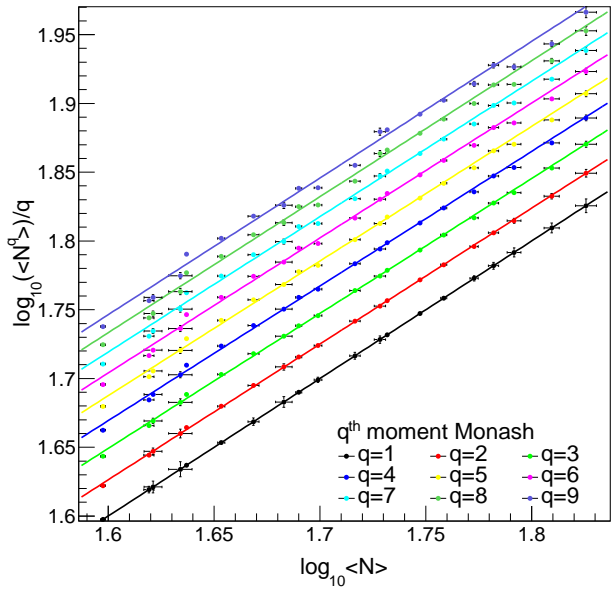}%
	\begin{center}
		\caption{\label{fig:moments} The first nine moments of \textsc{Pythia} 8 multiplicity distributions in function of the average multiplicity corresponding to each \pTjet window, normalized with their order $q$ on a log--log scale with linear fits.}
	\end{center}
\end{figure}

Figure~\ref{fig:moments} shows the values of each of the $q^{\rm th}$  moment up to $q=9$ in function of the mean, divided by their order, on a log--log scale.
In case Eq.~(\ref{eq:lambda}) is fulfilled, a linear fit with unity slope should describe each moment. While we are going to quantify it later, it is apparent from the figure that a linear description is generally adequate for all the moments in the whole \pTjet range.

\begin{figure*}[t]
	\centering
	\includegraphics[width=0.5\textwidth,keepaspectratio]{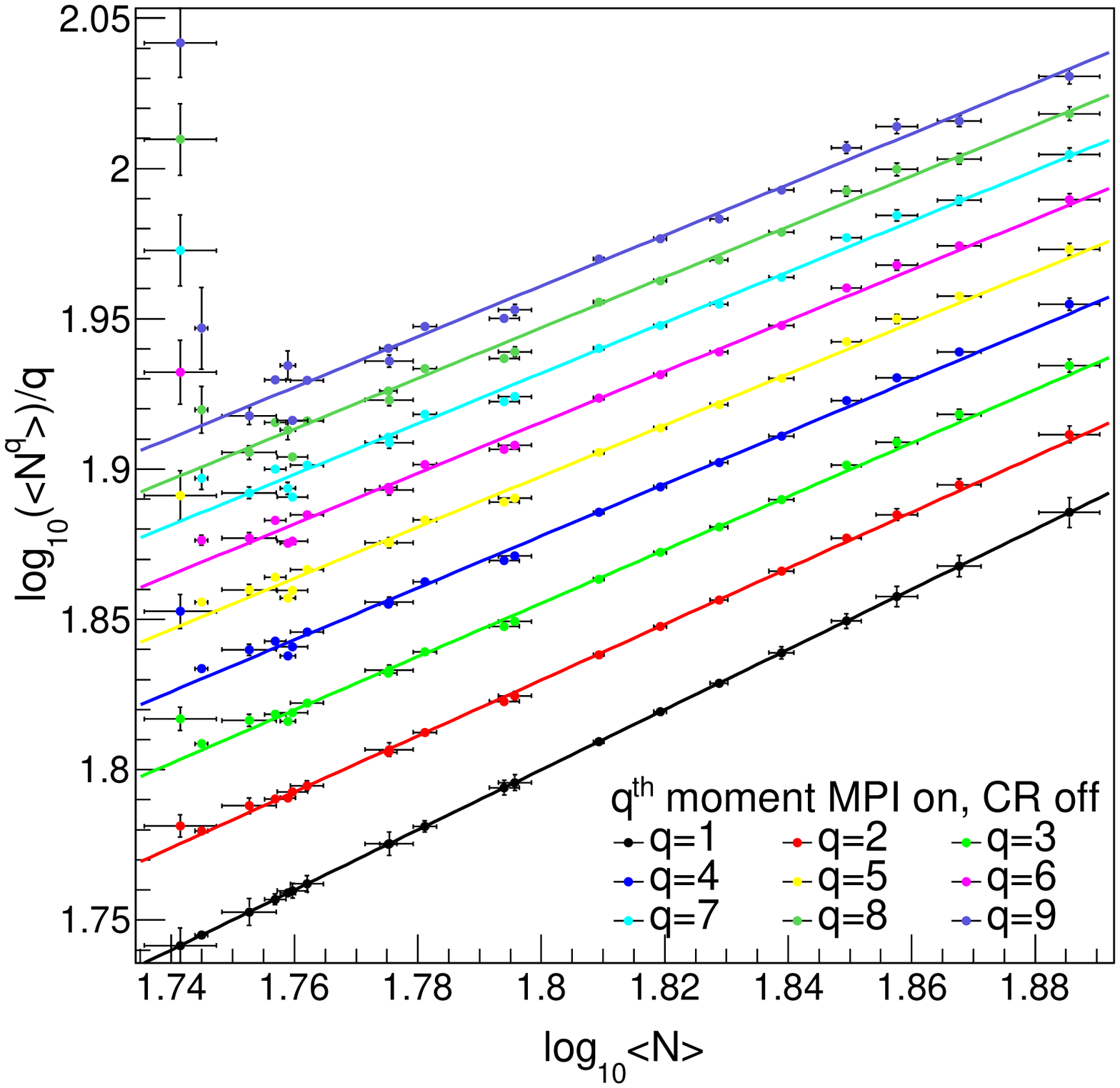}%
	\includegraphics[width=0.5\textwidth,keepaspectratio]{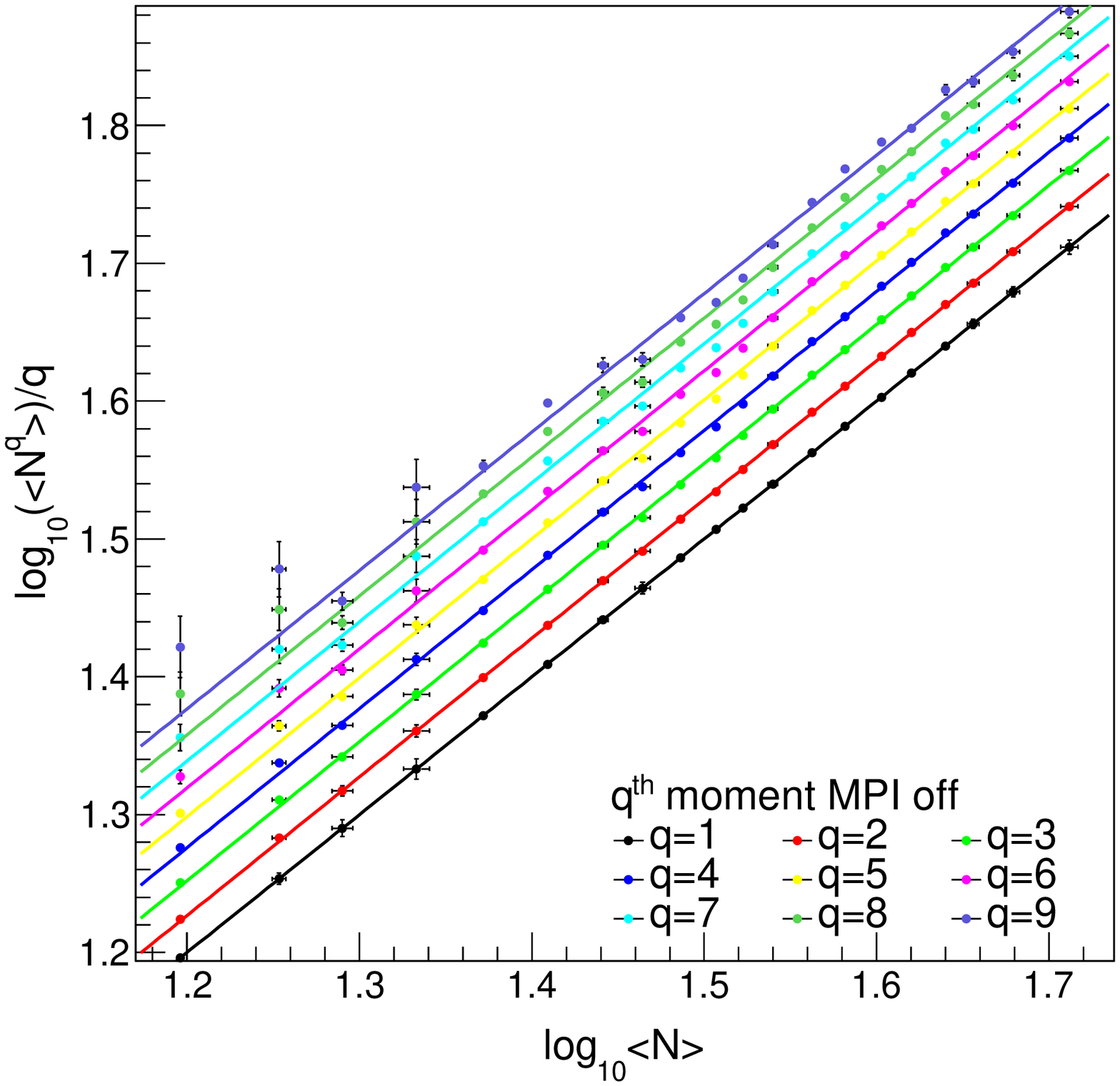}%
	\begin{center}
		\caption{\label{fig:momentsNonphys} The first nine moments of \textsc{Pythia} 8 multiplicity distributions without CR (left) and without MPI (right), normalized with their order $q$, in function of the average multiplicity corresponding to each \pTjet window, on a log--log scale with linear fits.}
	\end{center}
\end{figure*}

In order to understand the role of multiple-parton interactions with color reconnection in the emerging scaling, we repeat the above procedure with either CR or MPI turned off. The left panel of Fig.~\ref{fig:momentsNonphys} shows the results from simulations without CR, while the right panel presents results when MPI is turned off as well.
It is immediately apparent that the scaling is not fulfilled if the CR is off, as the slopes differ from moment to moment (the lines are not parallel). However, the scaling is restored if also the MPI is turned off.

\begin{figure}[t]
	\centering
	\includegraphics[width=0.5\textwidth,keepaspectratio]{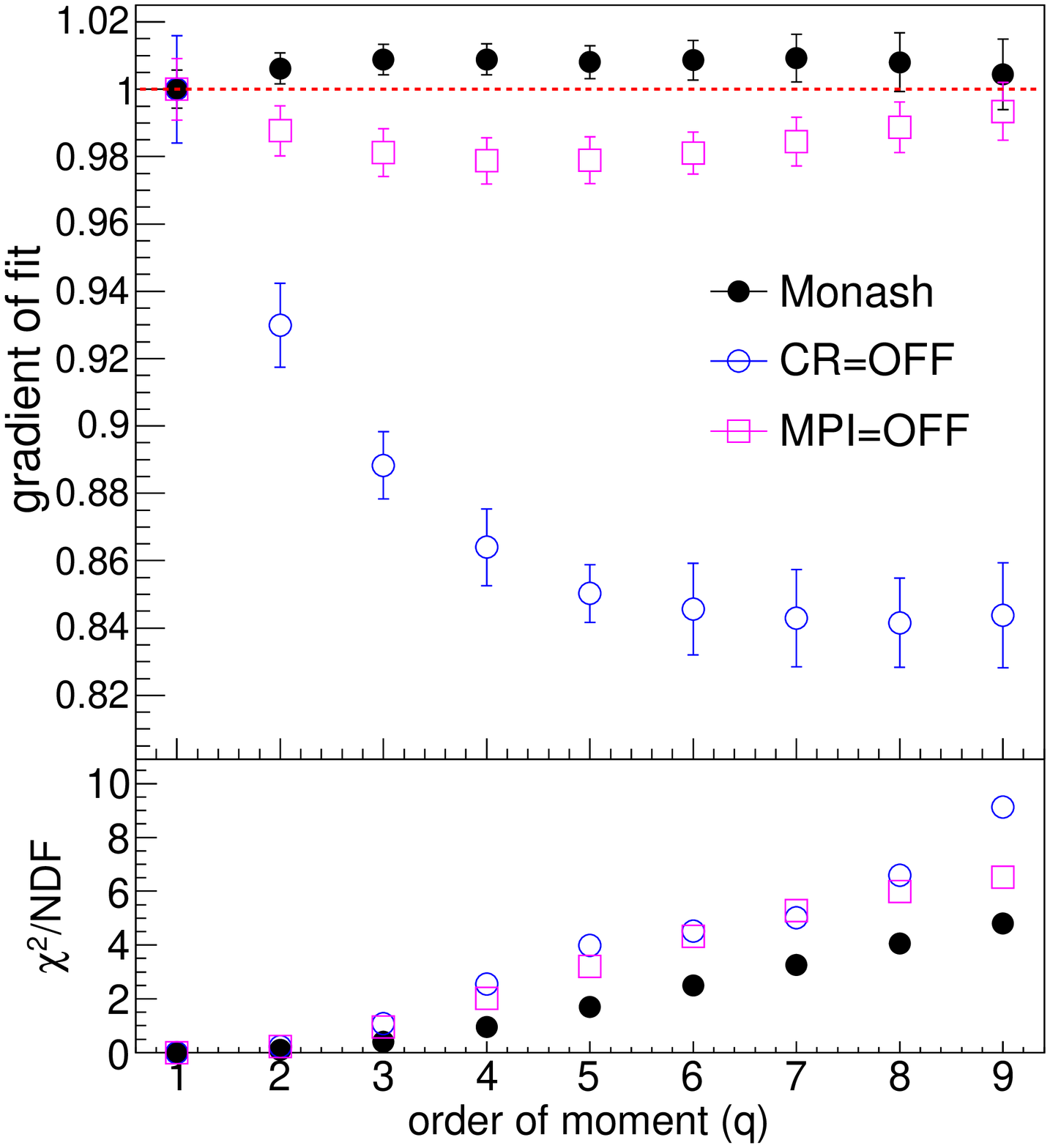}%
	\begin{center}
		\caption{\label{fig:Gradients} The gradients (top panel) and the $\chi^2/{\it NDF}$  of the fits (bottom panel), corresponding to each order of momentum.}
	\end{center}
\end{figure}

The above observations are quantified in Fig.~\ref{fig:Gradients}. 
Its left panel shows the goodness of the fits ($\chi^2/{\it NDF}$) for the physical simulation settings as well as for the scenario without MPI and without CR. The values are generally small, slowly increasing with the order of the momentum $q$, demonstrating that the $N^q$ assumption is statistically acceptable in all three cases. The fits are slightly better in the physical case than with the other two settings. The slopes of the linear fits are shown in the right panel of Fig.~\ref{fig:Gradients}. In the physical case all the moments are consistent with unity within 1\%, mostly within the range of the fit uncertainties. In the no-CR case, the discrepancy is as large as 15\% for higher moments. In the no-MPI case, the scaling is fulfilled with the accuracy of 2\%.
 
Events without MPI can be considered as physical events that are preselected so that no multiple-parton interactions occur. The development of color strings during the partonic processes is an inherent part of QCD. In \textsc{Pythia} 8, however, color flow is handled in a simplified manner, except for the hardest MPI, and CR is applied in a following step~\cite{Sjostrand:2017cdm}. Final states of events without CR are therefore not realistic. This explains why the scaling observed in physical events holds up in a no-MPI scenario, but breaks down when the MPI is present but CR is turned off.

{\it\label{sec:concl}Conclusions. --- }
We carried out a study of the multiplicity distributions of events with hard jets in proton-proton collisions at the LHC energy of $\sqrt{s}=7$~TeV using \textsc{Pythia} 8 Monte Carlo simulations. We demonstrated that the charged-hadron multiplicity distributions scale with the momentum of the jets. We also found that the scaling is held up in events without multiple-parton interactions (MPI), but breaks down in case MPI is present but color reconnection (CR) is turned off. Considering that the multiplicity-scaling within a jet is retained both in the no-MPI case and in the physical case (with MPI including CR), it is straightforward to assume that the KNO scaling is violated when the one-to-one connection is broken by complex QCD processes outside the jet development, such as single and double parton scatterings linking different hard processes within one event, as well as softer MPI with the beam remnants, similarly to the scenario proposed in Ref.~\cite{Ortiz:2018vgc}, or in the parton shower by overlapping color strings as suggested in Ref.~\cite{Abramovsky:1980yc}. Scaling of the jet radial profiles with event multiplicity is retained in the case where CR is off~\cite{Varga:2018isd,Gemes:2020cfa}. Therefore we conclude that the two scaling behaviors are not directly linked. While \textsc{Pythia} 8 describes event multiplicy distributions well and therefore we expect our findings to hold in real data, cross-checks with LHC measurements would rule out any dependency on the chosen model components. Requiring the scaling behavior of multiplicity with jet momentum can then serve as an important element in the development of new models.

{\it\label{sec:concl}Acknowledgement. --- }
The authors would like to thank S\'andor Hegyi for the discussions that provided important input to this research. This work was supported by the Hungarian National Research, Development and Innovation Office (NKFIH) under the contract numbers OTKA FK131979 and K135515, the NKFIH grants 2019-2.1.11-T\'ET-2019-00078, 2019-2.1.11-T\'ET-2019-00050, 2019-2.1.6-NEMZ\_KI-2019-00011 and 2020-1.2.1-GYAK-2020-00013, as well as the THOR Cost Action CA15213. The authors also acknowledge the computational resources provided by the Wigner GPU Laboratory and research infrastructure provided by the E\"otv\"os Lor\'and Research Network (ELKH).


\end{document}